# New scenarios and trends in non-traditional laboratories from 2000 to 2020

*Ricardo M. Fernandez, Member IEEE*, *Felix Garcia-Loro, Member IEEE, Gustavo Alves, Member IEEE*, Africa López-Rey, *Member IEEE*, Russ Meier, Fellow IEEE, Manuel Castro, Fellow IEEE.

**Abstract—For educational institutions in STEM areas, the provision of practical learning scenarios is, traditionally, a major concern. In the 21st century, the explosion of ICTs, as well as the universalization of low-cost hardware, have allowed the proliferation of technical solutions for any field; in the case of experimentation, encouraging the emergence and proliferation of non-traditional experimentation platforms. This movement has resulted in enriched practical environments, with wider adaptability for both students and teachers. In this paper, the evolution of scholar production has been analyzed at the global level from 2000 to 2020. Current and emerging experimentation scenarios have been identified, specifying the scope and boundaries between them.**

## I. Introduction

Labwork practices can be identified as educational activities carried out in a controlled environment, where there are controlled variables and observed ones. There is no doubt about the distinctive learning possibilities available in such environments. In a study carried out by the U.S. National Research Council, laboratory experiences were defined as "opportunities for students to interact directly with the material world (or with data drawn from the material world), using the tools, data collection techniques, models, and theories of science" [1, p. 31].

The goals and outcomes intended for laboratory activities may be diverse. Beyond the theoretical background, labwork includes a variety of transversal competencies such as: teamwork; gathering, analyzing, and interpreting data; computational thinking; communication; practical skills; promoting mastery in the study area; skills development for advanced knowledge; scientific reasoning; and empirical capacities [1]-[6]. Labwork also integrates easily into different approaches like: Project-Based Learning, Problem-Based Learning, Learning by Doing, Enquiry–Based Learning or Discovery-Based Learning, among others [3], [6]-[10]. Therefore, labwork is considered a key element for educational institutions when designing and constructing curricular content [1]-[4], [11]-[16].

Providing laboratory experiences does not guarantee students achieve the potential benefits offered by transversal competencies. Teachers may provide laboratories according to their needs or importance but fail to justify their presence for meaningful learning [17]. Therefore, integration of laboratories into curricular programs as well as the decision of making lab sessions mandatory or optional should be analyzed in the light of a global strategy to avoid a highly inefficient learning scenario [2], [11], [18].

In today's educational landscape, the provision of labwork environments adapted to educational methodologies, as well as to the needs of institutions, teachers, and students, is a priority. This need has encouraged the creation of solutions based on the adaptation of laboratory experiences to emerging educational environments [19]-[31]. The scientific community agrees on two main dimensions for distinguishing laboratory types: "relative location between the experiment and the experimenter, and the physical nature of the experiment" from which four main scenarios are well documented [19], [21], [22], [24]-[28], [32]-[34]. Based on the technological development of recent years, new alternatives have emerged, complementing and enriching laboratory solutions [29], [32], [33], [35]-[40]. Comparison among these scenarios is a recurring concern [20], [41]-[46]. However, each experimentation environment has advantages, as well as certain disadvantages over the others, either for users, institutions, teachers, or learning objectives. A direct comparison between the different alternatives is not possible due to the lack of a uniform criterion to evaluate them [20], [43], [46]. We can conclude that they have become complementary tools: the best solution is a combination of them [13], [43], [44].

In this article we address the evolution of these experimental learning environments. The research questions addressed are:
RQ1 What are the lab scenarios used in education? And what are the scope delimitations of each scenario?
RQ2 What are the types of emerging labs scenarios?
RQ3 How have lab scenarios evolved in the 2011-2020 decade?

## II. Methodology

The first step of the methodological process was collection of the core set of articles that address the topic. To obtain the corpus of scientific literature, scientific databases were consulted. The most important and widely used bibliographic scientific databases are Web of Science (WOS) and Scopus [47]-[49]. The Scopus database, produced by the ELSEVIER

.

o

group, includes a larger catalog of scientific production from journals and international conference proceedings than the Web of Science., An initial analysis showed that three out of five papers documenting non-traditional laboratories were published as conference papers. Therefore, the Scopus database was the preferred database in this research.

For the construction of the database query used to establish the corpus, we first identified the existing laboratory scenarios. To identify the dimensions that make up the different experimentation scenarios, we started with the classification that gathers the greatest consensus in the scientific community. As previously mentioned, this classification uses two main dimensions and results in four basic types of scenarios [19], [21], [22], [24]-[28], [32]-[34]. From this basic classification with four possible scenarios, the emergence of new scenarios that propose an alternative approach in the laboratory architecture were analyzed, showing hybrid solutions that combine these basic scenarios [24]-[29], [32], [33], [49]-[53], the proposal of completely alternative scenarios to the existing ones [16], [27], [38]-[40], [54], [55]. or the application of different approaches that entail substantial changes in the main scenario [23], [29], [31], [32], [56]-[60]. We analyzed how each of these labels is used to define the adopted laboratory solution. In no case was the use of the same solution in different environments considered a different scenario or dimension. For example, the use of a remotely accessible system used by student in the classroom or outside the classroom. Although both situations can provide different benefits, from the system point of view, the architecture is the same and we cannot consider it a different scenario or dimension.

Once the spectrum of laboratory scenarios were identified, we proceeded to the construction of a taxonomy that encompasses all the scenarios, and the labels used to delimit each environment. In this way, we built a common framework for the rest of the article.

To obtain the definitive query an iterative process was carried out, adding new keywords to the ones included in each previous query.

Based on the framework labels a query was built, new keywords related to the topic emerged, were analyzed and identified. These new keywords were processed and, if they seemed to be relevant for the research, they were added into the query. Likewise, their scope and scenario have been analyzed in the research that used the term. In this way, not only were new terms or labels identified, but also new scenarios. The goal was to identify unique scenarios for laboratories, and which variables come into play. After the identification stage, the definition of these scenarios was done through a label or name, as well as their delimitation, so each laboratory subtype was defined and delimited without ambiguities. Subsequently, we proceeded to group the terms and keywords that turned out to be synonyms for the identified scenario under the same label.

Considering that the term 'LAB' encompasses the terms "lab", and "labs", several iterations led to the debugged query shown below. The query does not use the '*' wildcard in a term such as "lab*" to avoid the appearance of terms such as "label", "labor", or "labour". Final query in which the terms related to "LABORATOR*" and "EXPERIMENT*" (replaced by ellipsis) have been removed as they imply replicated elements of the query of term "LAB":

TITLE-ABS-KEY ("cloud LAB" or "computer aided LAB" or "computer assisted LAB" or "computer based virtual LAB" or "concurrent LAB" or "cyber enabled LAB" or "cyber LAB" or "dataset based LAB" or "deferred LAB" or "digital LAB" or "distance LAB" or "distant LAB " or "distant LAB" or "distributed learning LAB" or "distributed virtual LAB" or "electronic LAB' or "hands on distance LAB" or "home LAB" or "homeLAB" or "hybrid LAB" or "hybrid local LAB" or "hybrid online LAB" or "hybrid reality LAB" or "hybrid remote LAB" or "internet assisted hands on LAB" or "internet assisted LAB" or "internet based LAB" or "LAB at distance " or "LAB at home" or "LAB kit" or "massive open online LAB" or "mobile LAB" or "mobile open online LAB" or "mono user virtual LAB " or "mono user virtual LAB" or "multi user virtual LAB" or "on line LAB" or "online LAB" or "pocket LAB" or "remote controlled LAB" or "remote hands on" or "remote LAB" or "simulated based LAB" or "simulated LAB" or "simulation based LAB" or "simulation LAB" or "smart LAB" or "software based LAB" or "software LAB" or "take home LAB" or "tele LAB" or "tele operated LAB" or "teleLAB" or "teleoperated LAB" or "ultra-concurrent LAB" or "ultraconcurrent LAB" or "virtual LAB" or "virtual manipulative LAB" or "web based LAB" or "web LAB" or "webLAB" or ….. or "web accessible simulat*" or "vlab" or "virtual training software " or "virtual reality mock up" or "virtual instrument" or "rlab" or "remotelab" or "remote simulat*" or "olab" or "o lab" or "mool" or "ilab" or "elab" or "e lab" or "dataset based simulat*") AND PUBYEAR > 2000 AND PUBYEAR < 2021 AND TITLE-ABS-KEY (education* OR training OR learning OR teaching OR instruction OR tutoring OR pedagog* OR tutoring) AND PUBYEAR > 2000 AND PUBYEAR < 2021

The scientific literature corpus was collected using the query, and 13,321 documents were obtained. Out of those documents, pre-processing was carried out, searching for errors, as well as the standardization of terms and the identification of erroneous terms. To obtain the consistent number of eligible articles, the PRISMA flow diagram shown in Figure 1 was used [61].

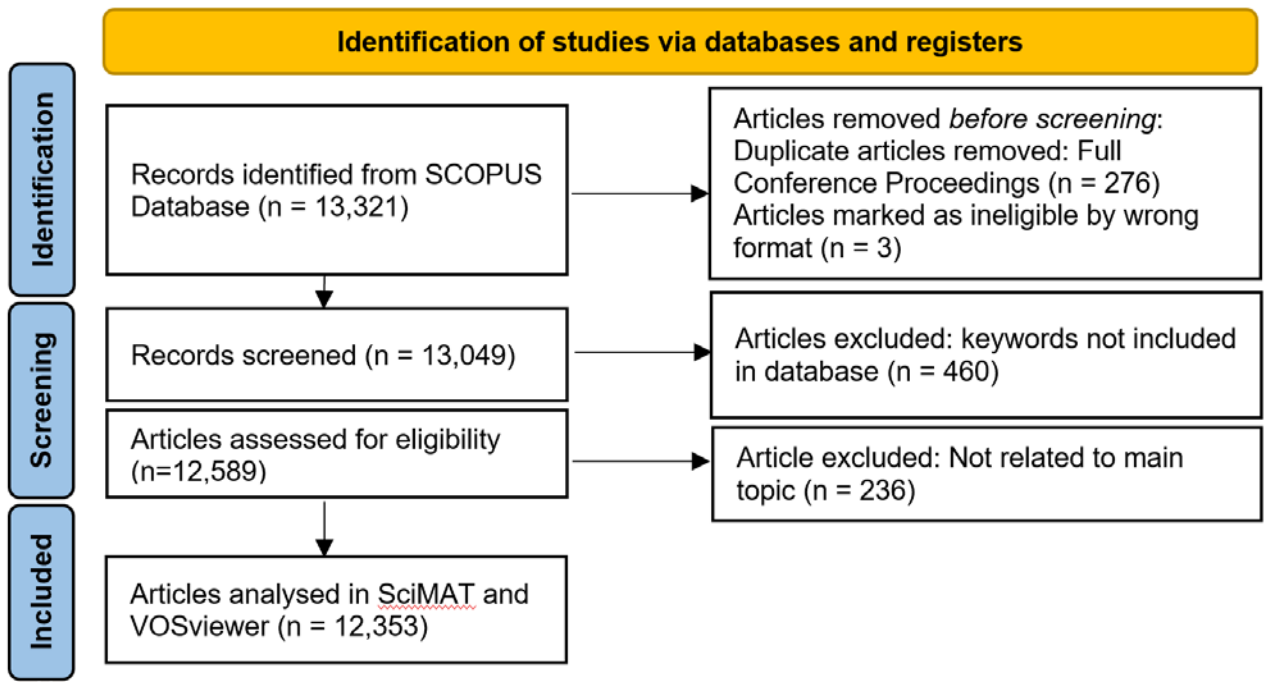

Fig. 1. Prisma flow diagram (literature selection)

Finally, we proceeded to the homogenization and grouping of terms. As an example, the keyword "Hybrid lab" includes all the following keywords: "hybrid laboratory", "hybrid laboratories", "hybrid labs", "hybrid laboratories in engineering" and, "hybrid remote laboratory".

Once the data has been preprocessed, there are multiple software tools to carry out this task: Bibexcel, Biblioshiny, CiteSpace, CitNetExplorer, SciMAT, Sci2 Tool, VOSviewer [62] as well as those provided by SCOPUS. SCIMAT software was used for the analysis. SCIMAT is an open-source (GPLv3) software tool created by the SECABA research group at Granada University [63]. SCIMAT software provides density and centrality graphs by periods, characterizing and classifying

o

the clusters in four groups, mapping them into a two-dimensional space (Figure 2). In the representation, centrality indicates the importance of the subject under study within the field, and density represents the degree of development of the subject studied [48], where centrality is defined as: $c=10\times\sum e_{kh}$, with $k$ an item linking to a topic and $h$ an item linking to other topics. Density measures the internal degree of interrelations in the network and is calculated as follows: $d=100(\sum e_{ij}/n)$ with $i$ and $j$ items belonging to the theme and n being the number of items in the topic. The coefficient $e_{ab}=(c_{ab}^2)/(c_a \times c_b)$ is called equivalent index, where $c_a$ is the number of words of $a$, $c_b$ is the number of words of $b$ and $c_{ab}$ is the number of co-occurrence of words $a$ and $b$.

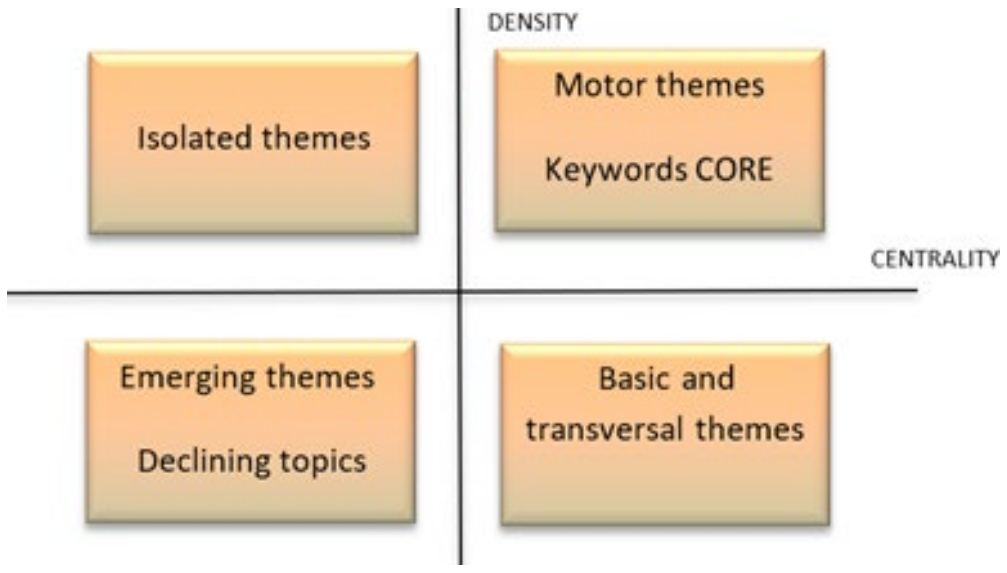

Fig. 2. Four-quadrant representation

Keywords in the upper-right quadrant, with strong centrality and high density, include the core themes. These themes are externally related to concepts applicable to other themes, which are conceptually closely related. The themes in the upper-left quadrant include isolated keywords; themes in the lower-left quadrant consider disappearing themes or themes that may emerge in the future. Finally, the last quadrant (lower-right) refers to transversal and basic themes, that could become core themes in the future.

## III. Identification of the Dimensions in the Laboratory Operation Scenarios

Unfortunately, there is no classification unanimously supported by the specialized community, beyond those that leave out emerging and alternative scenarios. These discrepancies arise in the dimensions used to classify the different types of labs. A basic classification is the one-dimensional one used by [20], which divides laboratories into two large groups: traditional labs to refer to those corresponding to on-site physical laboratories within an institution, and non-traditional labs to refer to those which are accessible remotely, usually through the Internet and whose nature may be based on real systems, or simulated ones. The grouping of remotely accessible labs, regardless of their physical nature, is recurrent in numerous studies under different terminologies, such as online labs, digital labs, electronic labs (e-labs), web-based labs, distance labs or virtual labs, as shown in Table 1.

However, as aforementioned, there is a solid consensus of two core dimensions: relative location between the experimenter and the experiment – local or remote, and the nature of the experiment – real or virtual. The development of new technologies and the application of new methodologies have led to the emergence of new solutions that require a redesign of this model. The names given to these types of laboratories vary from author to author. These differences in typology will be detailed later, to identify the terms to be used and define the scope of each laboratory.

Hybrid solutions emerge from core scenarios. Several authors contemplate hybrid solutions using the nature of the experiment "real/simulated" both locally and remotely [24]-[27], [29], [32] [50]-[53], whilst other authors have extended this approach and developed hybrid systems based on a real device locally manipulated together with a distant real laboratory, or a simulated one [24]-[26], [28], [29]. The model proposed by [29] includes all the hybrid approaches based on core dimensions found in the literature.

In a hands-on laboratory, the equipment can be manipulated directly, in a traditional way, or using a computer-based interface (e.g., a LabVIEW VI). Based on this idea, in [21] the authors add a third dimension, based on the type of interaction aforementioned: controlling the devices and instrumentation equipment directly or through a computer-based interface. Strictly speaking, it cannot be considered a third dimension: in a lab based on a computer model the interaction can only be carried out through a computer-based interface. Remotely accessible scenarios require interactive means. In other words, this third dimension exists only in experiments whose nature is real and locally interacting.

In [54] the authors consider a third dimension, analyzing if the lab environment is used by one or by multiple institutions. However, in this research we do not consider this characteristic, since it is a property of the laboratory administration system, either through an ad-hoc solution, or through a Remote Laboratory Management System (RMLS), and, therefore, independent of the laboratory.

Some proposals only affect one type of laboratory scenario, leading to subtypes. A dimension that only makes sense within real systems which are remotely accessible is the one provided in [30]. In [30] and [69] authors discriminate between two types of interaction: interactive laboratories and batch laboratories. Interactive laboratories refer to those whose connection and interaction are continuous during the experimentation, while in batch laboratories the access is queue based. In the same scenario (remote and real), authors in [39] consider that there are two types of laboratories: those whose interaction occurs in real-time (or quasi-real time) and those in which all possible interaction with the physical laboratory has been stored in a way that is based on actual results recorded in a database. Based on this premise, authors in [37] go further and propose a scenario by combining both laboratory subtypes in a hybrid way.

Within remotely accessible solutions, the terms massive open online labs (MOOLs) [70]-[74] and Mobile Massive Open Online Laboratories (M-MOOLs) [75] have been mentioned. For the authors, the term MOOL only makes sense in remotely accessible physical systems since concurrent access is a representative characteristic of computer-modeled systems.

Authors in [76] identify two types of laboratory operation depending on the interaction possibilities within the experiment

o

(parameters tunability) and in the experimentation scenario (changeable experiment environment) for the experimenter: fixed/open laboratory. Authors in [24] adds a third type of experiment or laboratory: Observation laboratory. For their part, [77] add a fourth scenario, with the possibility of guided semi-open laboratory. These types of operation modes are more likely to be provided in some types of laboratories. However, they are not exclusive to any of them and can even coexist allowing high flexibility in the laboratory at different learning levels.

An attempt to classify laboratories, providing them unambiguous names and their potential relations is described in [29]. In it, the authors define a notation based on the Unified Modeling Language (UML). They consider laboratories divided into two main dimensions: Physical and Online ones, with hybrid combinations. However, their classification suffers from a space dedicated to local simulated labs.

A disruptive classification is the one offered by [77]. In this research, the authors use three dimensions: pedagogical approach, degree of virtualization and laboratory distribution. The degree of virtualization encompasses both the nature of the experiment and the distance between experiment and experimenter.

However, its dimension in terms of the "laboratory distribution" referred to the local/international access to the lab and is meaningless in face-to-face laboratories, as well as in remotely accessible laboratories since the location of the experimenter is not an influential variable for the laboratory. In short, we do not consider this dimension as valid to identify an exploitation scenario.

Laboratory classification is deeply analyzed in the next section, the different alternatives are contemplated, considering core dimensions, hybrid alternatives, and emerging scenarios, as follows:

- *Labs based on real systems:* Experimentation in real environments whose occurrence is in real-time. This dimension includes local and remote scenarios.
- *Labs based in Software:* Experimentation based on modeled systems. This dimension includes local and remote scenarios.
- *Deferred Labs:* Laboratories based on real environments, where events are deferred over time. The laboratory provides data previously stored in databases for this purpose.
- *Hybrid Labs:* A fusion or combination of different scenarios.

## IV. Type of Laboratories

The first drawback that has been detected is the use of terms without a formal definition that establishes the physical framework in which the laboratory is deployed, and the limits in which the experimentation takes place [64], [65], [78]-[81]. This ambiguity is further complicated by the diversity of terms used to identify the same experimentation environment, or the use of the same term to define different environments, as already highlighted by [13], [28], and [56]. Still many authors establish a formal definition or description of the laboratory [13], [19], [37], [82]-[90], but in many cases these are not consistent definitions with the real scenario [91], [92], partially not consistent [24], [25], [93]-[96] or conflictive with the definitions provided in other studies. The classification of the terms found in the literature is collected in Table 1, which includes most of the bibliography used for this paper. The focus of the table is the classification of diverse designations used by researchers in the existing body of literature, according to the identified scenarios in this paper, which is considered relevant and useful for further analysis on the subject.

### *A. Labs based on real systems.*

Traditional laboratories are based on physical systems based on specific instrumentation, equipment, physical environments, and real components/elements. Real-scenario experiments give students unique experiences such as facing errors, uncertainties, failures, and/or unexpected results in the real physical world [29], [31]. Within physical laboratories, they can be divided into two options, according to the relative location between experiment and experimenter: local laboratories and remote laboratories. Local laboratories are developed in scenarios in which all the elements of the laboratory and the experimenter are in the same location, and in which the experimenter can observe and measure on the physical system, either directly or by means of instrumentation. Remote laboratories are physical systems for experimentation in which the experimenter accesses and/or observes and/or remotely controls the experiment through a network, generally the Internet.

#### *1) Local laboratory interaction*

Traditional centralized physical labs are the classic laboratories in academic institutions. Scalability has been solved by replicating workbenches. In recent years, new educational resources have emerged that have led to the development of lower-cost practical activities by using inexpensive components. Additionally, they are also used in conjunction with open or low-cost software. The resulting laboratory allows students to perform simple laboratory experiments like those performed in a traditional and more expensive laboratory. Additionally, users can experiment at home or elsewhere, without time restrictions. These physical distributed laboratories are commonly defined/named as mobile laboratories.

The authors in [32] argue that the mobile laboratory concept refers to having low-cost hardware so that students can create their laboratories. In [31], authors mention that "If electric or electronic components are used, a limitation will be for standard and cheap components, as any other solution will not be feasible due to financial reasons." In the literature, there are different synonyms and definitions for these laboratories (see Table 1).

From an educational point of view, pocket labs focus on self-study development, although they have also gained much ground in the academic environment of universities and colleges.

Pocket laboratories are a very useful tool in the first years of engineering careers, where there is a high component of theoretical and abstract learning, bringing students closer to the practical activities that help cement the theoretical concepts they are acquiring [97]. In this approach, the role of the instructor

o

| Scenario | Label | Terms Found in Scientific Literature | |
|---|---|---|---|
| **Real Local- Centralized**<br>Real Lab<br>Centralised Lab<br>Centralised User | Hands-on Lab | Classic Lab [31]<br>Computer Aided/Assisted Experiment [156] [157] [209]<br>Computer Aided/Assisted Lab [154] [155] [208]<br>Face to Face Lab [56]<br>Hands-on Lab [20] [21] [24] [30] [32] [33] [42] [44] [45] [54] [97] [109] [117] [118] [132] [143] [150] [151] [153] [154] [209]<br>In-Person Lab [56] | Local Lab [25] [152]<br>On-Site Lab [29] [110]<br>Physical Lab [105] [118] [153]<br>Physical Manipulative Experiment (PME) [49] [151] [158] [159]<br>Real Experimentation [83]<br>Real Lab [41] [43]<br>Traditional Lab [20] [26] [34] [77] [109] [152] |
| **Real Local- Distributed**<br>Real Lab<br>Distributed Lab<br>Distributed User | Pocket Lab | At-Home Lab [169]<br>Distance Lab [36] [56] [57] [58] [166]<br>Hands-On Distance-Lab [166]<br>Home Experimental/Experiment Kit [36] [60] [170]<br>Homelab [23] [59] [162]<br>Individual Experimental Kits [143] | Lab at Home [163]-[165]<br>Mobile Lab [29] [32] [110]<br>Mobile/Massive Open Online Lab (MOOL) [32]<br>Pocket Lab [31] [35] [97] [98] [160] [161]<br>Remote Lab [171]<br>Take-Home Lab [166]-[168] |
| **Hybrid (Remote & Pocket)**<br>Real Lab<br>Centralised & Distributed Lab<br>Distributed User | Hybrid | Hybrid Lab [29] [32] [33]<br>Remote Haptic Lab [99] | |
| **Real Remote**<br>Real Lab<br>Centralised Lab<br>Distributed User | Remote Lab | Cyber-Enabled Lab [23]<br>Distant Lab [86] [183] [184]<br>Distributed Learning Lab [182]<br>Hands-on Remote Lab [211]<br>ilab [175]<br>Internet-Assisted Hands-on Lab [196]<br>Internet-Assisted Lab [196]<br>Lab at Distance (LAD) [190]-[192]<br>Massive Online Lab (MOOC Lab) [142]<br>Massive Open Online Lab (MOOL) [70]-[74] [143]<br>Online Lab (OLAB) [23] [175]<br>Real Virtual Lab [128]<br>Remote-Controlled Lab [68] [133]<br>Remote Hands-on Lab [45] [177] | Remote Lab (RLAB) [5] [7] [13] [19]-[21] [23]-[26] [28]-[34] [37] [41] [43] [44] [46] [51]-[54] [56] [69] [76] [77] [79] [81] [82] [84] [86] [88]-[90] [99] [104]-[119] [125] [132] [136] [141]-[147] [150] [152] [172]-[174]<br>Remote Web-based Lab [7]<br>Remotelab [193] [194] [195]<br>Telelab [46] [185] [186]<br>Tele-Operated Lab [187]-[189]<br>Virtual Instruments [64]<br>Virtual Lab [64]-[68] [176]<br>Web-Based Lab [178]-[180]<br>Web-Based Remote Lab [90]<br>Weblab (Web Lab) [175] [181] [210] |
| **Simulated Local - Distributed**<br>Simulated Lab<br>Distributed Lab<br>Distributed User | Lab Simulator | Computer-Based Virtual Lab [123]<br>Computer-Aided/Assisted Lab [153] [198] [199]<br>Computer-Simulated Lab [87]<br>Local Simulation [21] [24]-[26] [32] [33] [110] [152]<br>Multi-User Virtual Lab [19] [84] [22] | Simulator [30] [37] [44] [121]<br>Simulation Lab [49] [118] [172]<br>Software-Based Lab [197] [200]<br>Virtual Lab [49] [78] [87] [118] [125] [150] [153] [197]<br>Virtual Manipulative Experiment (PME) [49] [151] [158] [159] |
| **Simulated Remote**<br>Simulated Lab<br>Centralised Lab<br>Distributed User | Virtual Lab | Computer Aided Lab [204]<br>Cyber Lab [125]<br>Digital Lab [203]<br>Distributed Learning Labs [182]<br>Distributed Virtual Lab [150]<br>Internet Based Lab [201]<br>Mono-User Virtual Lab [19] [22] [84]<br>Online Virtual Lab [207]<br>Online Lab (OLAB) [24] [56] [152]<br>Remote Simulation [21]<br>Simulation-Based Lab [77] | Simulated Lab [5] [13] [109]<br>Simulation Lab [23] [54] [172]<br>Virtual Lab (VLAB) [13] [19] [20] [22]-[26] [28]-[30] [32]-[34] [41] [43] [46] [51]-[54] [84] [85] [90] [99] [108] [109]-[111] [115] [118] [119] [129] [132] [146] [152] [173] [176] [197] [207]<br>Virtual Lab Simulation [80]<br>Virtual Manipulative Lab [158] [202]<br>Web-Accessible Simulation [205] [206]<br>Web-Based Lab [207]<br>Web-Based Virtual Lab [44] [90] [207] |
| **Deferred Labs**<br>DDBB Lab<br>Centralized Lab<br>Distributed User | Deferred Lab | Dataset-Based Simulator [37]<br>Deferred Lab [38] [39] [40]<br>Ultra-Concurrent Lab [16] [39] [55] [54] [136] | Virtual Lab [135]<br>Virtual Reality Mock-Up [54] |
| **Hybrid (Remote & Deferred)**<br>Remote & Deferred Lab<br>Centralized Lab<br>Distributed User | Hybrid | Hybrid Dataset and Reality Lab [37] | |
| **Hybrid (Remote & Virtual)**<br>Real & Simulated Lab<br>Centralized Lab<br>Distributed User | Hybrid | Hybrid Lab [24] [29] [30] [32] [33] [52] [53] [110] [146] [147] [152]<br>Hybrid Online Lab [25] [27] [50] [51] [152]<br>Hybrid Remote Lab [26]<br>Online Hybrid Lab [52] [53] | |
| **Hybrid (Pocket & Simulator)**<br>Real & Simulated Lab<br>Distributed Lab<br>Distributed User | Hybrid | Haptic Virtual Manipulative Lab [49]<br>Hybrid Lab [25] [28] [29] [152]<br>Hybrid Local Lab [26]<br>Virtual Haptic Lab [99] | |

Table. 1. Laboratory scenarios

as facilitator, besides guiding and motivating students, generates and designs contents for learning [98]. The elements and components used in the development kits must be easily accessible and reasonably priced [31]. It is intended that all components are reused for new experiences, thus reducing wast of materials so that the activity is ecologically sustainable. While commercial products are one solution, there are also open-source hardware and software platforms that give instructors a powerful tool to arouse curiosity and learning in students. These have become widely used in the last years and allow students to create their laboratory at home with low-cost boards and sensors for the development of practical activities.

Pocket labs can be turned into a hybrid system if the mobile lab is an element of a complex system forming a symbiotic relationship between the local laboratory (mobile lab) and either local/remote software or remote hardware (see Table 1). Several combinations are possible [99], however some of them have not been reported yet.

For some years now, the DIY (do it yourself) movement has gained relevance [100]. This approach seeks to promote lifelong learning by expanding digital skills, autonomy, and creativity of

o

students through collaborative learning. Authors in [100, p. 7] highlight that "in recent times, it is not only attracting a significant number of science enthusiasts and talented lay people within and beyond the borders of universities and traditional research institutions but has also caught the public imagination as well as the attention of governments, policymakers and academics." Another philosophy based on DIY movements and open-source software is what is known as a Fabrication Laboratory (FabLab) [101]. FabLabs are a production space with physical hardware, equipment and instruments which incorporate electronic control and software. This philosophy promotes placing the experimenter at the center of the training experience, where they themselves produce the materials and learning activities. In [102], the authors defines it as follows: "A DIY laboratory is a place, set up by interested person(s) or group, equipped for scientific experiments, research, or teaching in which numerous private and community-based initiatives use scientific methods alongside other forms of inquiry such as hacking and remixing to engage with techno-scientific concerns and societal challenges". In this sense, Ferreti [102, p. 1] argues that "The locations of DIY science are various and range from small groups of tech enthusiasts to large online communities that share scientific objectives and outcomes, organized into local makerspaces, Fabrication Laboratory (Fablabs), classrooms, universities and museums, public libraries and private enterprises."

*2) Physical remote lab*

Complex experimentation with physical equipment in real environments is often costly in terms of time, money and qualified staff as it requires the development and maintenance of learning infrastructures, kept in good condition [26], [34], [77] In addition, it is frequent that, once these infrastructures are deployed, they remain underused, fundamentally due to the limitation in the available time to be used by students.

The rapid advancement of technology and the exponential use of the Internet in education have enabled practical scientific research to be performed in physical laboratories accessible through the Internet. Accessing and controlling physical laboratory devices through the Internet presents a solution to solving the learning needs of students, members of the workforce, and scientists by providing access to expensive and complex equipment remotely. In this regard, these labs are an appropriate tool to provide 24/7 availability to accessing a physical lab, in all educational levels and purposes, as well as in research and training in industry.

The term widely used to name this experimentation environment is "Remote Laboratory.", Authors use this term to describe a remote laboratory with physical equipment at some distance from the user who accesses the experimentation through the Internet [27], [81], [90], [104]-[112]. In this sense, the authors in [20] identify remote labs as "(…) being characterized by mediated reality. Similar to hands-on labs, they require space and devices." This same idea is highlighted by [21], [24], [25], [37]. The difference from hands-on labs is the distance between the experiment and the experimenter [20].

Another relevant topic in remote labs are the elements from which it is made up. A remote laboratory is a tool composed of software and hardware layers that manage the access, validate the accessing request, and allowing users to remotely access real equipment, located in the university through the Internet. Along the same lines, [89] [113], [114] emphasize the existence of two large systems within laboratories based on remotely accessible physical systems: the software-based block and the laboratory based block. The software-based block oversees the communications (human-machine, machine-machine) and interface between human and laboratory. It also manages several features such as authentication, visualization in some cases, request management, and user management, . The laboratory-based block consists of the equipment, instruments, elements, environment, and devices required for experimentation. In [115], the authors provide a comprehensive definition about the scope of remote laboratories where they say: "Remote Labs use real plants and physical devices which can be used at a distance, either by teleoperating and receiving measurements from them or by modifying some input parameters and allowing simple visual observation." This definition is consistent with a broad scope of authors [12], [26], [29], [31], [51], [53], [54] [113], [116]. In this line, the authors in [117] provide a similar definition, however they emphasize remote labs as an educational resource.

*3) Real remote and local hybrid lab*

These hybrid laboratories are based on having local hardware that allows learners to control remote hardware through the Internet. This type of laboratory allows the user a physical manipulation of the experiment and a way to teleoperate a real piece of equipment at a distance [29], [32], [33], [99]. Through this approach, manual skills, the main drawback of online labs, are provided during laboratory experimentation. However, under this laboratory variant, the existing bibliography is based on the development of theoretical models, without even the development of proofs of concept [29], [32], [33] when compared against real systems [99].

### *B. Lab based on software.*

Laboratories based solely on software solutions have fields of application that laboratories based on real systems cannot reach. These are experiments where implementation would require extraordinarily complex means, experimentation about unobservable phenomena, destructive tests, not be scalable, experimentation would involve living beings, be long in time and/or expensive [12], [118], [119]. Examples can be found in the building sector [120]; chemical processes [80]; botany [121]; agroforestry [122], biology [123], [124]; pharmaceutical [124] or forming technology [81]. The authors in [85] highlight the similarities between laboratories based on physical systems and those based on models, which have common features, view and control, but are separated attending to the nature of the experimentation environment, the "model" used.

For these laboratories whose existence is limited to software, there is an evident tendency to use the term "virtual", especially to simulated laboratories which are remotely accessible. However, the term virtual lab is not used in a homogeneous way when identifying and defining the scope of the labs (see Table 1). This lack of consistency in terminology and in its formal definition between studies was already pointed out long ago by

o

[13], [125], and it continues to be a recurring problem [23] [28], [56].

*1) Simulated local lab*

Within laboratories based solely on software solutions, types can be clearly identified where the software resides on the machine from which the user interacts, so their use can even occur offline. In the different studies, the terms most used for this type of laboratory involves the concept of simulation in some of its forms and with less weight that of virtual, although both terms are used in a similar way to define an environment of laboratory that does not exist physically. At this point, the difference between simulator and simulated laboratory must be highlighted: while a simulated laboratory aims to model a real laboratory and the operations carried out in it, a simulator is not concerned with the experimentation environment and only with the results.

*2) Simulated remote lab*

The term used most often for simulated laboratories in different studies is, without a doubt, "virtual lab". In [87], the authors establish the basis of the contemporary definition of virtual laboratory when they state that a "virtual laboratory is a simulation of a physical laboratory" and that "Virtual laboratories are computer-simulated laboratories that look, operate, and produce results similar to real laboratories." Therefore, they reserve the term "virtual laboratory" for referring to those laboratories whose environments are based on simulated processes regardless of the locations of experiment and experimenter. This definition is consistent with the ones used by [13], [82]-[84], [118], [120], [126] where authors discriminate two types of environments: simulated plants and real plants. Along the same lines, the authors in [93] establish a definition based on simulations and computerized models that can be reached both locally (distributed virtual laboratories) and remotely (centralized virtual laboratories). However, they include in the concept of virtual laboratory supplementary activities to the laboratory or to experimentation such as forums, videos, and glossaries. [94], [126], [127].

Authors like [86], [87], [128] stress on a key aspect related virtual labs: the similarity of the virtual experimental environment with the hands-on experimental environment. In this regard, [129] focuses on similarity in controls and operations while [130] highlights that the output data must be indistinguishable from data from a 'real' experiment. For its part, [83] in its definition emphasizes the interactivity of the simulation with [118] highlighting that a requirement for the virtual laboratories is that "student must feel like they are working with real authentic devices in a real authentic space". From this requirement, they formulate four criteria that cover the aspects highlighted above in terms of the similarity of the environment, results, behaviour, and control. But they add one more component: "laboratory space must be created which allows for communication and collaboration among students and with the lab supervisor." In this way, [118] reference the social and collaborative factor of traditional face-to-face labs.

On the other hand, many authors use the term virtual to refer to the way in which interaction with the laboratory is carried out [66], [67]. In this way, any laboratory in which the interaction is carried out through a software-based interface, even if the system on which the orders and/or control are executed is a real system, is defined as a virtual laboratory. Hatherly [68] provides a definition that encompasses the scenarios: "A virtual laboratory is one where the student interacts with an experiment or activity which is intrinsically remote from the student, or which has no immediate physical reality" and highlights the interaction channel by saying: "via a computer interface". This is the case of the system presented by [65] in which the term virtual laboratory and remote experimentation are used indistinctly to refer to a real laboratory accessible and controllable through the Internet (remote laboratory). Similarly, [91] emphasizes the concept of virtual through the use of software to cover the physical distance between the user and the real system. In this way, they define a virtual laboratory as a tool that allows "an emulation or simulation allowing distance interaction with the laboratory instruments".

One confusing and recurring term in the different studies is that of "Virtual Instrument," widely used in practice settings [129]. Virtual Instrument uses the term virtual to refer to the interface of an instrument built using software that can be made up of controls (knobs, buttons, etc.) and indicators (graphs, LEDs, screens, etc.) [131]. This Virtual Instrument can be connected to real instruments, for example through communication standards (e.g., GPIB, PXI, VXI, RS232, or RS485) [132], [133], or be based in a modelled system [129], [132], [133]. Therefore, the term can be ambiguous to define a laboratory since a Virtual Instrument does not establish if there is a real physical system or if it is a simulated environment for obtaining the results. Other contexts in which the term virtual has been used in the laboratory environment do not conform to the character used in this article. In 1967, [134] describes an experimental program for learning math which transforms the traditional arrangement of the classroom. Authors labelled it as "virtual laboratory for learning". However, students manipulate physical materials. Obviously, in 1967 the virtual term was used outside of the scope of the contemporary term, which is based on computerized processes. Also [88] uses the term Synchronous Virtual Laboratory to describe a remote laboratory service: "to follow online a laboratory activity held by the teacher." Although the authors consistently describe a distributed remote laboratory scenario, the term virtual does not respond to the described scenario; however, the term is intended to differentiate a series of services provided to students.

Despite the diversity, there is a high consensus in the establishment of the term "virtual laboratory" as those laboratories whose environment and results are based solely on software [12], [13], [20], [84]. It is considered a solid experimental model and an exclusively software-based environment. However, there are more characteristic aspects in virtual laboratories [13], [20], [87]. Therefore, in this article we use the term virtual laboratory to refer to those laboratories whose environment is based on software, whose results are obtained through pre-established algorithms and whose control and behaviour mimics a real system. Within virtual laboratories, regarding the relative locations between the virtual laboratory and the end user, in [12] the authors use the terms "virtual mono-user lab "to classify virtual laboratories whose interaction takes place locally with the "user interacting on the machine where

o

the laboratory software is located", and "multi-user virtual lab" to refer to those virtual laboratories whose interaction is carried out remotely where "the laboratory is located on a different machine from the user, so that it accesses the laboratory through some type of network." In this study we use the term virtual lab to refer to those software-based laboratories accessible through the Internet, while the term simulated lab is used for software-based laboratories locally accessible.

### C. *Deferred Labs*

The most recent technological addition to this pool of laboratories is based on considering the potential combinations of the actuators and sensors in a physical laboratory, recording in real-time all the potential results, values and videos and storing all this information for later use [135], [136]. In this way, user-laboratory interaction reproduces the results of a real environment without the inconveniences of scalability, breakdowns, and maintenance of a real laboratory, although both the experiment measurements and viewing are not happening at the precise moment of the interaction.

A similar approach to this strategy has been traditionally used in virtual reality mock-ups for, among other objectives, training reasons [137]-[140]. For these reasons, the authors consider deferred labs as a scenario included in scenarios with remote access, and due to their characteristics, it cannot be classified as either real, virtual, or hybrid.

The first record of this type of laboratory resource appears on Stanford's online experimentation iLabs platform [141]. They do not provide a formal name for this laboratory paradigm. They describe it as digital versions of real-world science experiments, a virtual laboratory environment obtained from a physical experiment and based on images and data logged into a database, in which students interact with the same controls and video view as would be used during a real experiment session but observing prerecorded states. This description is consistent with the one provided by [39] as "a laboratory based on a set of pre-recorded experiences carried out at a real lab" as well as with the ones found in [37], [38], [40], [136]. However [54] and [136] highlight the recording of a certain number of copies, for example, for each possible state of the laboratory / experiment so that the end user does not always obtain the same results in the same state of the experiment, but, as in a real laboratory, there are small variations around the point equilibrium state of the experiment.

The process to store data of the potential experimental combinations and to obtain the consequent results and visualizations and test them may be automated [39], [55], [135] or not [39]. In fact, they can be developed from a hands-on traditional laboratory and a video camera, with later generation of the web-interface adding controls for actuators and screens for sensors measurements or visualizations.

To identify this type of laboratory, [38] highlights the way in which the experiment is carried out using the term "deferred laboratory", a term that [39] also recognizes. However [39] as well as [16], [40], [54], [55], [136] highlight the scalability aspect of the "ultra-concurrent lab." In [37], the authors highlight the storage aspect using the term Dataset-based simulator. However, the term ultra-concurrent lab may not delimit well the technological field of this type of laboratory and may be confused with those laboratories used in massive environments and known as Massive Online Laboratory [142] or Massive Open Online Laboratories (MOOL) [143]. In addition, the limitation in the number of simultaneous users is not only hardware but also software, so the line between the present scenario and those laboratories that allow concurrent access through batch mode [30] [143], [144], through the federation of the same laboratory in different institutions [30] [39], [113], [145] or by providing a redundant laboratory, [30] [39], [142], [143] can be very thin. For these reasons, in this article the term deferred laboratory is used since it reflects faithfully the execution time of the experimental process.

Despite the limited bibliography on the matter and the different names used, there is a high consensus regarding the definition and coverage of this type of laboratory. In conclusion, we can define it as a web laboratory in which all possible interactions have been previously obtained from a physical laboratory and which visualization, results, and measurements have been pre-recorded and stored in a database. In this article, we propose a name for this type of laboratory, deferred remote lab.

### D. *Hybrid Labs (Software and Hardware)*

The combination of simulations and real systems, both accessed locally or remotely, are commonly accepted as hybrid laboratories by the research community [24], [25], [27], [50], [51]. This combination is intended to provide a more complete experience for users by taking advantage of the benefits and main characteristics of both simulated and physical environments. Simulated and real environments allow users to quickly experience mathematical models based on software and experimentation in the real world where disturbances and failures typical of physical experimentation are generated that are not reflected in a software program or a simulation algorithm. Hybrid laboratories allow learners to have the main advantages of real laboratories and virtual laboratories based on software, regardless of the user's location. This allows complementary experiences through local simulation and then realization through the real experience. Or, the user can access a hybrid remote laboratory where they can first perform a software-based simulation in the same way and, for example, then remotely test this simulation on real hardware to contrast its operation [25], [52], [146]-[148]. Within hybrid laboratories where the benefits of both types of simulated and real laboratories are integrated, there are different possibilities according to the user's location, if they access the experiment locally or remotely, presenting the options of local hybrid labs [24]-[26], [28], [29], remote hybrid labs [24], [25], [27], [50], [51] and the third option of hybrid labs with remote hardware and deferred laboratory [37]. We can conclude that any combination of a simulator, real and deferred system, accessible remote, local and remote-local would be included in the hybrid laboratory term.

## V. Results

For the data analysis step of the research, the SCIMAT software was used following the workflow shown in Figure 3.

o

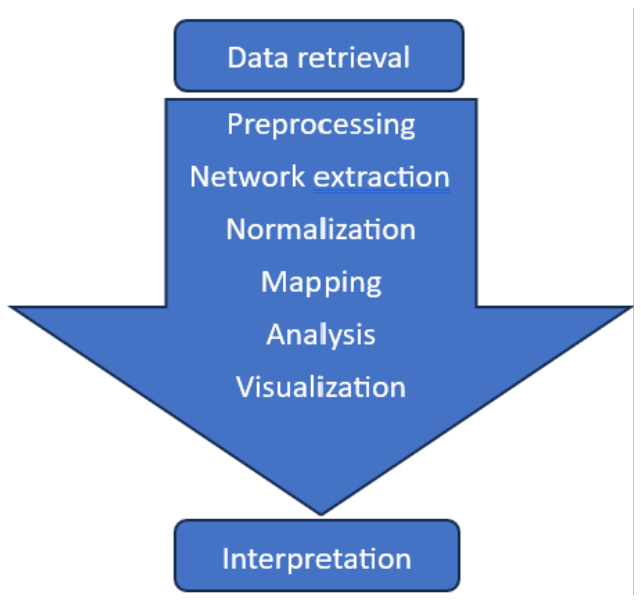

Fig. 3. Workflow of science mapping

In co-word analysis, clusters of keywords and their interconnections are obtained. These clusters are considered as themes [48] [63]. Each research theme (cluster) is characterized by two parameters: density and centrality. These parameters are used to classify themes into four groups, as shown earlier in Figure 2. Thus, a research field can be understood as a set of research themes, mapped onto a two-dimensional space [48]. Each cluster will consist of a minimum and maximum number of nodes or keywords. For example, in Figure 4, the cluster consists of six nodes (the upper limit established in the study). Each cluster is represented by a sphere, a number (389 in Figure 4), and a label in its corresponding period diagram (VIRTUAL-LAB in Figure 4). The volume of the sphere is proportional to the number of documents belonging to the cluster, the number represents the number of documents, and the label is assigned by selecting the central node of the cluster. Similarly, each node within the cluster is depicted in the same manner, with its size being directly proportional to the frequency of the represented keyword. Within the cluster, the thickness of the links between keywords is proportional to the equivalence index and represents the degree of association between both keywords. Regarding the value of a cluster, we can distinguish between "core documents" and "secondary documents." Core documents are those documents that have at least two keywords within the cluster, while secondary documents refer to those documents that have only one keyword within the cluster. The union of core documents and secondary documents is shown in the results provided in Figures 5 to 8.

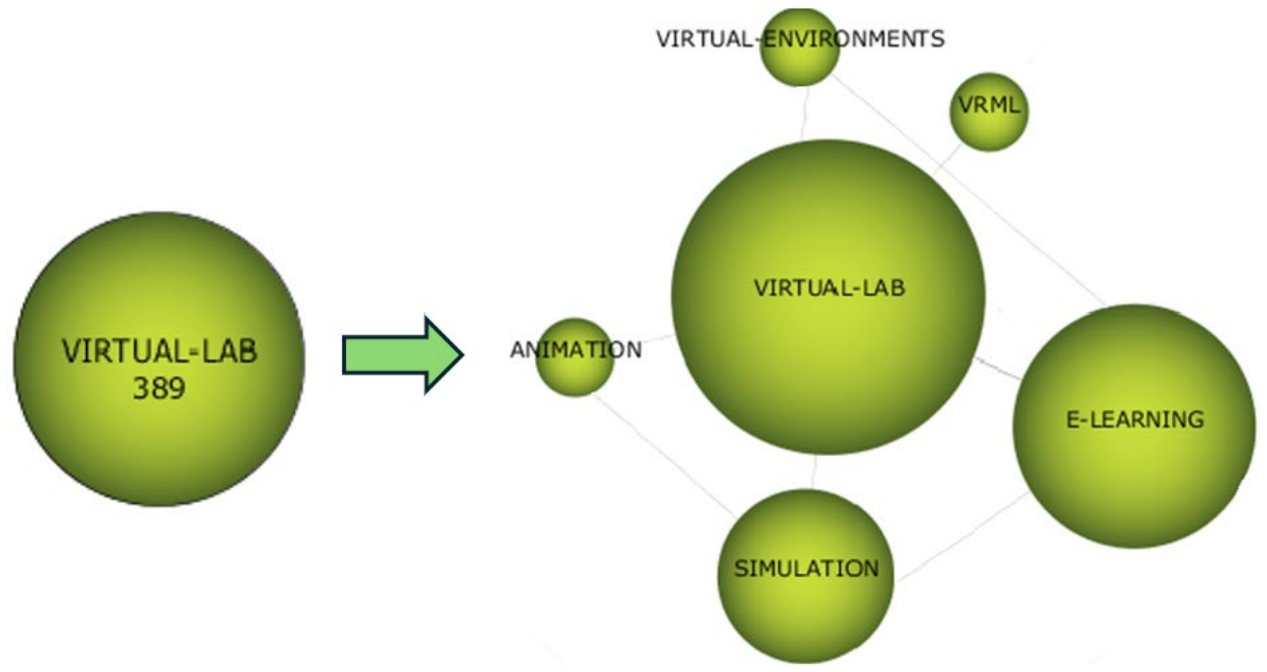

Fig. 4. Cluster network example.

Some parameters were set up, with the following criteria:

- *Type of analysis:* co-occurrence relations.

*Minimum co-occurrence:* 2 for two-year periods; 3 for decade periods.

- *Clustering Method:* to avoid irrelevant or weak associations simple center algorithm has been used.
- *Min frequency: 3,* only the item that appears in at least three documents in each period will be considered.
- *Cluster limits*: Max cluster size: 6; Min cluster size: 3.
- *Normalization:* equivalence index.

To begin the keyword analysis, a first study was generated covering the decade 2001-2010 (Figure 5) and then the decade 2011-2020 (Figure 6) to take a picture of the general situation and observe and compare the overall evolution of the core words. It is important to emphasize that over the analyzed years, the clusters' networks, as well as the keywords of each cluster, evolve and may change.

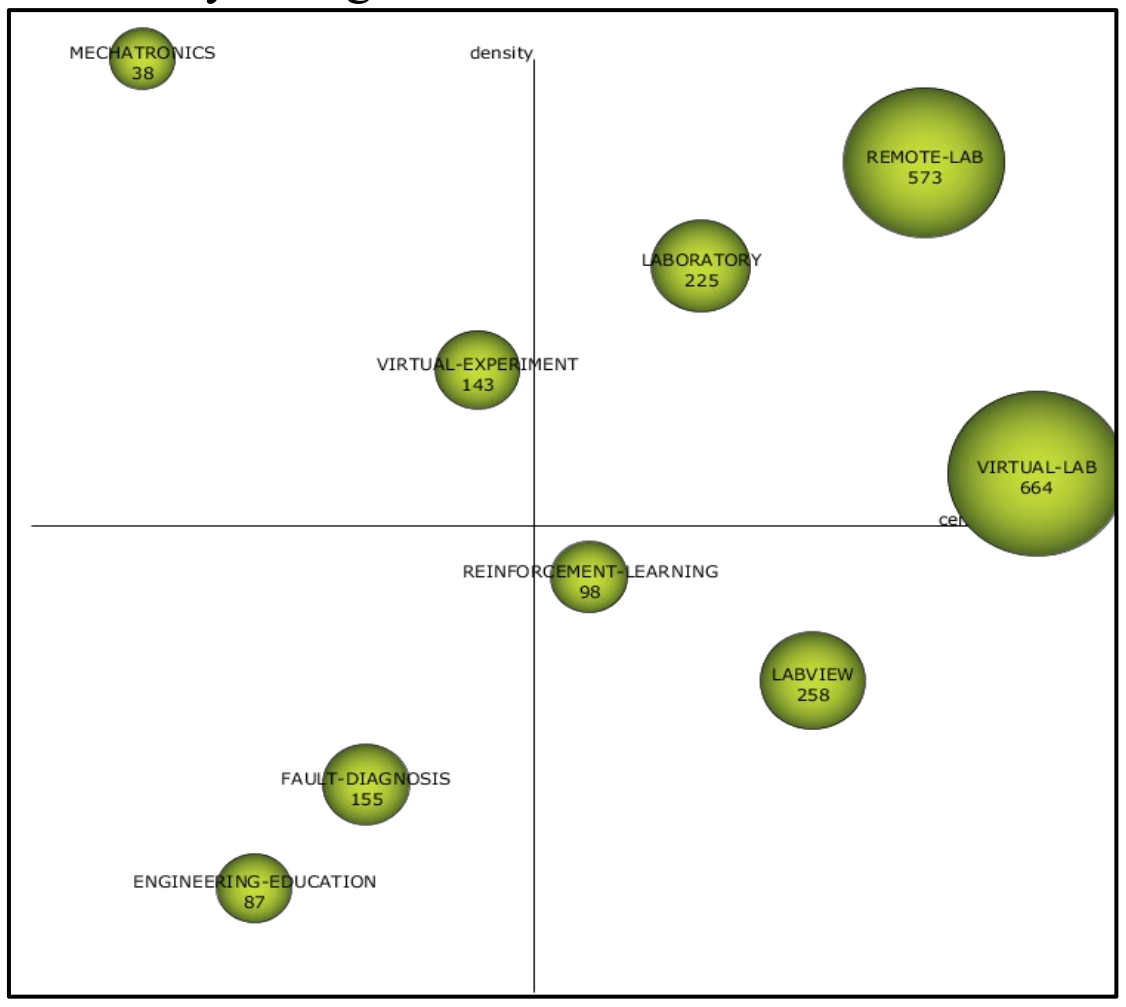

Fig. 5. Keyword analysis 2001-2010

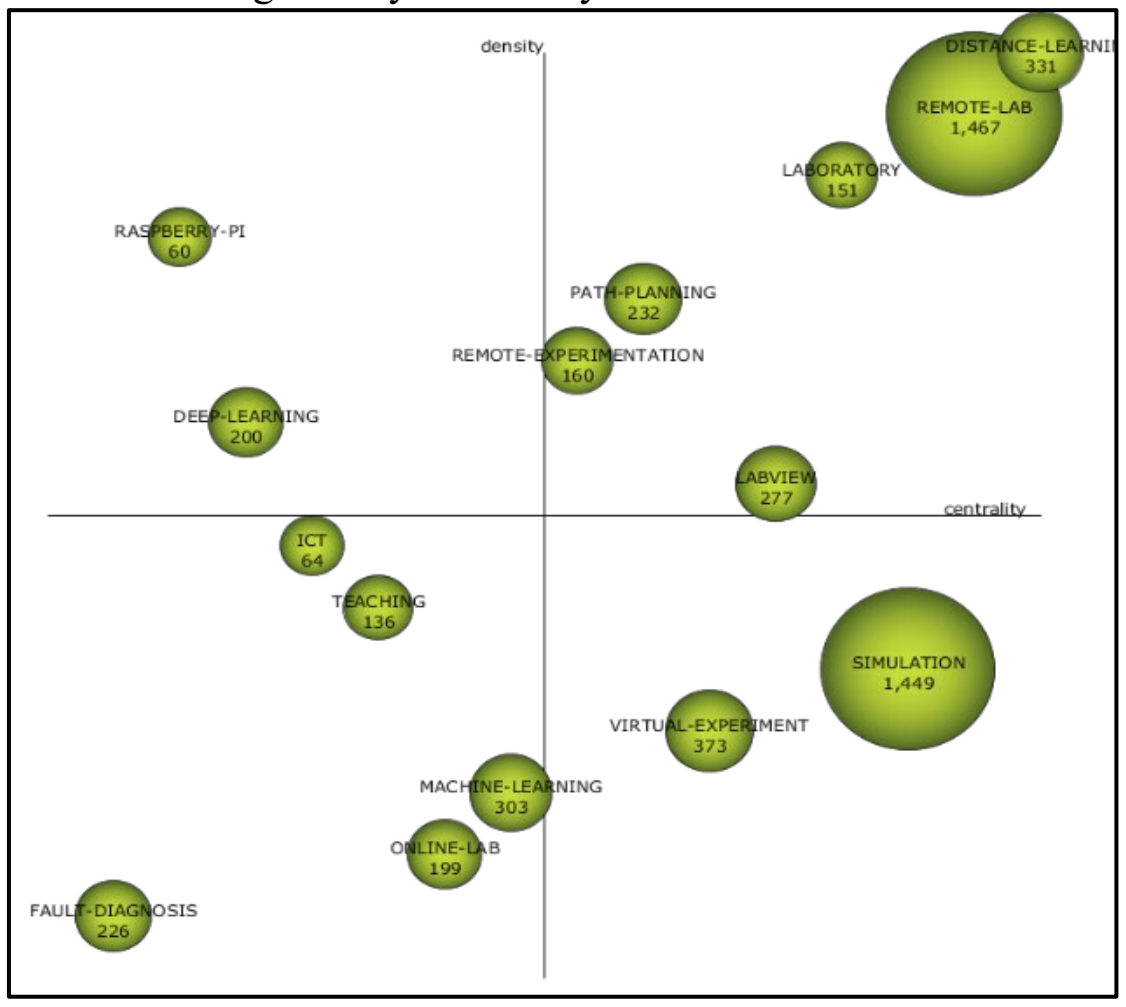

Fig. 6. Keyword analysis 2011-2020

In this first analysis by decade, the REMOTE LAB cluster is the main subject of the study in (2001-2010); in the second decade of the analysis, the number of appearances increases significantly to 1467 and it gains importance as a core theme.

The keyword VIRTUAL LAB with a strong presence in the decade (2001-2010) is in the SIMULATION cluster, in the second decade analyzed, and experiences a moderate movement between the two periods.

It should be noted that in the first period VIRTUAL LAB is the keyword with the highest number of documents in the bibliography, which are strongly related to education and simulation topics. The use of software and simulation tools has fueled the growth of virtual labs. Among the most popular software tools used in the first decade were LABVIEW and MATLAB, which maintained moderate growth in the second decade of the study.

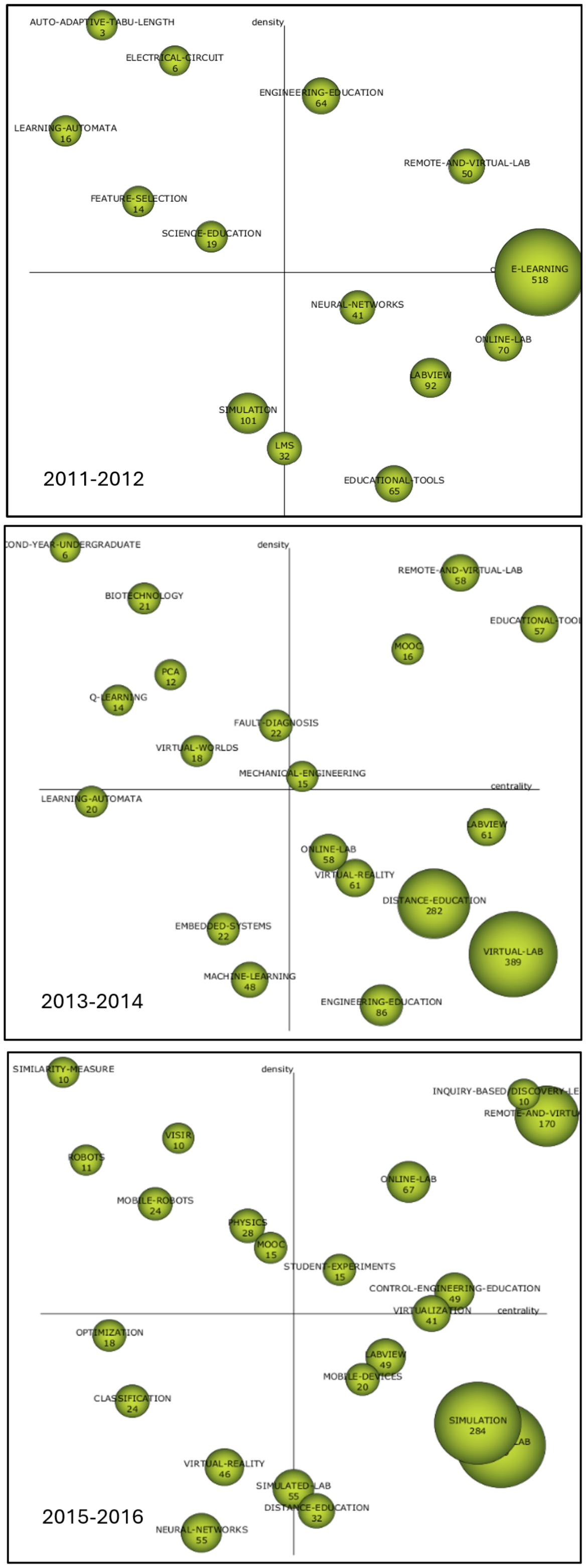

Fig. 7. (2011-2012), (2013-2014) and (2015-2016) Analysis

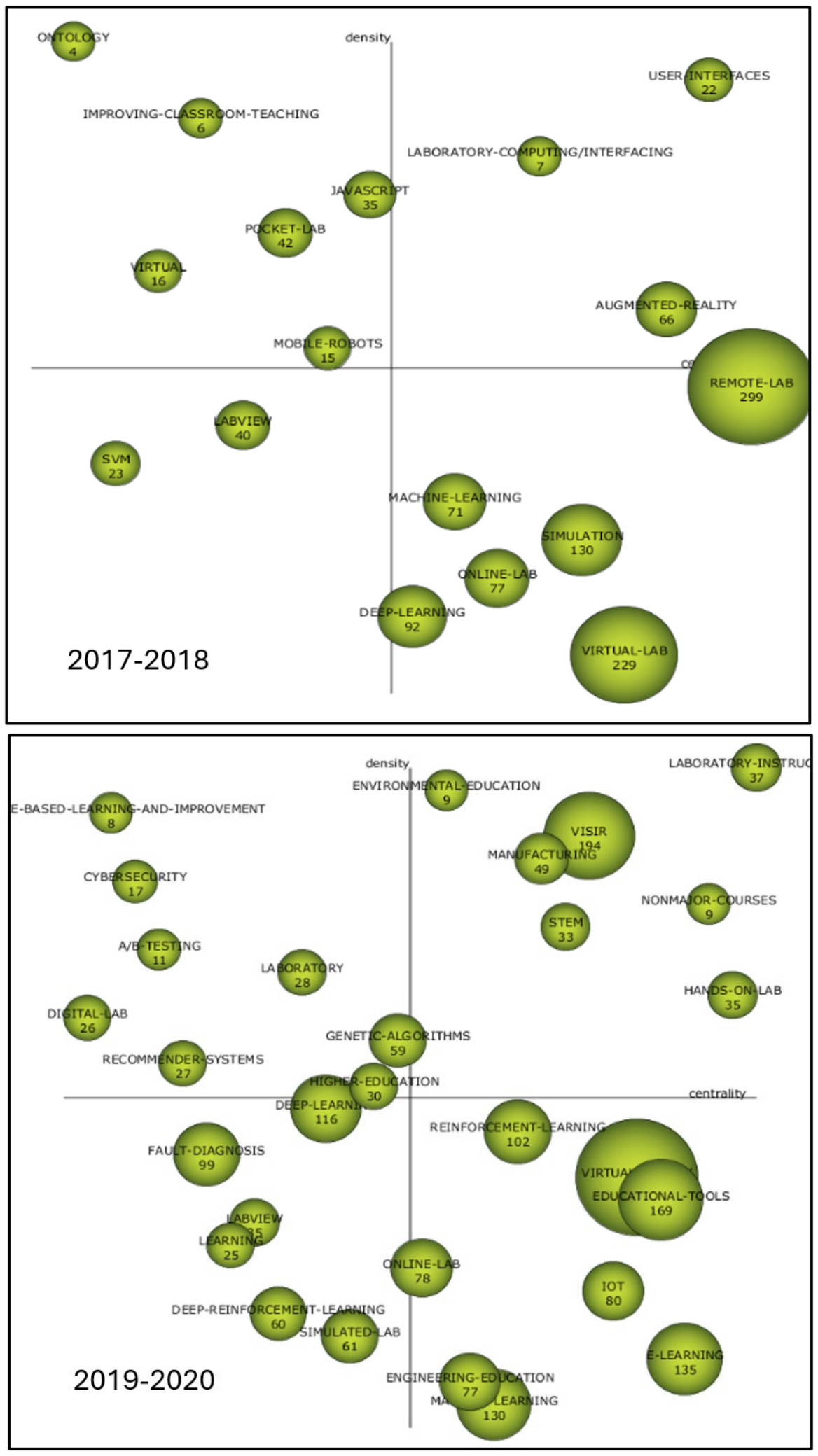

Fig. 8. (2017-2018) and (2019-2020) Analysis

In the case of remote labs, they are associated with the development of LMS in the first decade, with advances in the topic of web services as a starting point for remote experimentation.

It is also observed that the growth of the field of knowledge in Control Engineering Education encourages the use of non-traditional experimentation [84], as this is one of the most prolific areas in the studied period [116][132].

In the second decade, the sustained growth of open-source systems and the use of mobile devices for experimentation in learning environments should be noted.

After the first decade, a very strong growth in the term REMOTE LAB from 259 (first decade) to 952 (second decade) documents can be observed, while VIRTUAL LAB grows from 411 to 881 documents from the first to the second decade.
Since the goal is to observe and analyze trends, we will focus on the last decade (2011-2020) to deepen the analysis. In this decade, two-year periods were considered, and the results analyzed to assess the evolution of the main topics related to remote laboratories, virtual laboratories and topics that have become relevant in recent years. These are summarized in Figures 7 and 8.

Table 2 below provides a detailed list of the clusters belonging to each of the two-year periods studied, indicating for

| | Nº | CLUSTER | C | D | CD | UD |
|---|---|---|---|---|---|---|
| 2011-2012 | C1 | **remote-and-virtual-lab (24)**, haptic-devices (3), laboratory(16), web-based-design-tools (5), control-engineering-education (14), educational-laboratory (4) | 7.87 | 8.79 | 11 | 50 |
| | C2 | **labview (51)**, virtual-instrument (53), power-electronics (3), monitoring (3), digital-signal-processing (3), robot-control (3) | 5.1 | 3.13 | 22 | 92 |
| | C3 | **e-learning (156)**, virtual-lab (222), remote-lab (248), remote (8), distance-education (25), vocational-training (5) | 17.67 | 4.55 | 127 | 518 |
| | C4 | **engineering-education (56)**, learning-assessment (3), learning-technology (3), telelab (7), weblab (9), e-portfolios (3) | 2.51 | 13.36 | 12 | 64 |
| | C5 | **simulation (86)**, visualization (10), assessment (9), nursing-education (6), inquiry-learning (3), experiential-learning (5) | 1.86 | 2.69 | 15 | 101 |
| | C6 | **educational-tools (40)**, mobile-devices (4), open-sources (47), learning (15), curriculum (5), ict (5) | 4.3 | 1.57 | 8 | 65 |
| | C7 | **online-lab (40)**, online-course (8), web-services (13), computer-lab (7), rlms (16), university-education (3) | 10.4 | 3.65 | 13 | 70 |
| | C8 | **lms (20)**, blended-learning (7), evaluation (9) | 2.44 | 1.69 | 4 | 32 |
| 2013-2014 | C9 | **remote-and-virtual-lab (24)**, web-based-design-tools (6), laboratory (18), rapid-prototyping (3), physiological-systems (6), control-engineering-education (26) | 9.63 | 31.59 | 11 | 58 |
| | C10 | **labview (36)**, virtual (3), sound-card (3), motion-control (4), virtual-instrument (32), distance-lab (3) | 10.66 | 5.9 | 18 | 61 |
| | C11 | **distance-education (29)**, remote-lab (264), laas (6), lifelong-learning (9), online (4), vocational-education (3) | 8.76 | 3.91 | 27 | 282 |
| | C12 | **online-lab (42)**, personalization (3), network (8), remote-control (11), laboratory-learning (4), inquiry-learning (6) | 1.99 | 5.24 | 12 | 58 |
| | C13 | **virtual-reality(34)** , student-experiments (6) , augmented-reality (15), vrtools (3), web-based-lab (13), interactive-applications (3) | 6.37 | 4.79 | 11 | 61 |
| | C14 | **virtual-lab (214)**, simulation (91), e-learning (152), vrml (5), virtual-environments (6), animation (5) | 12.63 | 2.54 | 78 | 389 |
| | C15 | **engineering-education (54)**, hands-on-lab (6), educational-technology (11), m-learning (16), distributed-systems (5), evaluation (10) | 6.78 | 1.6 | 15 | 86 |
| | C16 | **mooc (7)**, visir (9), mool (4), distributed-learning (4) | 6.91 | 17.96 | 5 | 16 |
| | C17 | **virtual-worlds (8)**, second-life (3), simulated-lab (11) | 0.42 | 7.07 | 3 | 18 |
| 2015-2016 | C18 | **remote-and-virtual-lab (32)**, e-learning (137), labicom (4), ejs (5), web-based-design-tools (10) , laboratory (20) | 16.03 | 13.25 | 19 | 170 |
| | C19 | **online-lab (60)**, remote-sensing (3), inquiry-based-learning (8), undergraduate-education (3), biology (4), concept-maps (3) | 4.8 | 8.31 | 12 | 67 |
| | C20 | **simulation (102)**, virtual-lab (172), educational-tools (48), nursing-education (4), control (11), resuscitation (3) | 10.43 | 3.3 | 50 | 284 |
| | C21 | **labview (27)**, rlms (9), virtual-instrument (13), isa (4), sensors (7), pid-controllers (6) | 3.52 | 4.33 | 14 | 49 |
| | C22 | **remote-lab (266)**, lms (17), engineering-education (54), electronics (5) ,raspberry-pi (12), learning-analytics (13) | 15.19 | 3.18 | 62 | 299 |
| | C23 | **inquiry-based/discovery-learning (4)**, curriculum (5), laboratory-instruction (5), first-year-undergraduate/general (3), internet/web-based-learning (3) | 15.82 | 38.42 | 5 | 10 |
| | C24 | **virtual-reality (28)**, visualization (5), augmented-reality (13), gamification (6), haptics (3) | 0.77 | 2.49 | 9 | 46 |
| | C25 | **control-engineering-education (26)**, educational-technology (9), web-based-lab (11), remote-control (7), educational-laboratory (4) | 6.21 | 4.69 | 6 | 49 |
| | C26 | **simulated-lab (25)**, m-learning (14), hands-on-lab (19), interactive-simulations (3) | 2.33 | 1.83 | 5 | 55 |
| | C27 | **visir (7)**, lab-federation (4), comparative-evaluations (3) | 0.39 | 11.11 | 4 | 10 |
| | C28 | **mooc (11)**, oer (3), serius-games (5) | 1.71 | 6.46 | 4 | 15 |
| | C29 | **distance-education (19)**, control-engineering (6), computer-aided-instruction (11) | 2.34 | 1.81 | 4 | 32 |
| 2017-2018 | C30 | **user-interfaces (3)**, data-analysis (3) , biology (6), cloud-lab (4), learning-analytics (10), inquiry-based-learning (7) | 7.15 | 50.67 | 3 | 22 |
| | C31 | **augmented-reality (23)**, medical-education (3), virtual-reality (48), virtual-lerning-environments (3), mixed-reality (5), computer-vision (3) | 6.44 | 6.51 | 14 | 66 |
| | C32 | **javascript (9)**, lms (12), html5 (8), experimentation-language (4), experiments (11), java simulations (5) | 2.45 | 29.79 | 7 | 35 |
| | C33 | **remote-lab (209)** , mooc (29), e-learning (107), visir (22), embedded-systems (11) , engineering-education (30) | 16.36 | 6.04 | 92 | 299 |
| | C34 | **simulation (100)**, nursing-student (3), standardized-patient (3), modeling (5), nursing (4), educational-tools (35) | 4.9 | 1.72 | 18 | 130 |
| | C35 | **virtual-lab (214)**, ict (7), physics (10), inquiry-learning (7), conceptual-understanding (4), higher-education (7) | 4.96 | 0.9 | 20 | 229 |
| | C36 | **online-lab (48)**, stem (13), a/b testing (3), oer (7), assessment (7), learning (10) | 3.44 | 1.35 | 10 | 77 |
| | C37 | **pocket-lab (6)**, iot (27), teaching (7), raspberry-pi (7), electrical-engineering (6) | 2.09 | 14.23 | 6 | 42 |
| | C38 | **labview (24)**, ethernet (3), microcontrollers (3), measurements (3), virtual-instrument (18) | 0.94 | 4.07 | 9 | 40 |
| | C39 | **laboratory-computing/interfacing (3)**, laboratory-instruction (4), general-public (3), computer-based-learning (4) | 4.44 | 41.84 | 4 | 7 |
| | C40 | **virtual (5)**, online (3), laboratory (12) | 0.35 | 11.11 | 3 | 16 |
| 2019-2020 | C41 | **laboratory-instruction (23)**, distance-learning (23) , second-year-undergraduate (10), upper-division-undergraduate (9), first-year-undergraduate (19), computer-based-learning (14) | 52.37 | 48.37 | 29 | 37 |
| | C42 | **manufacturing (4)**, engineering (10), mooc (15), digital-learning (3), industry 4.0 (11), blended-learning (18) | 5.64 | 16.2 | 6 | 49 |
| | C43 | **visir (18)**, remote-lab (187), lab-federation (9), rlms (8), electronics (5), electricity (3) | 6.6 | 16.3 | 22 | 194 |
| | C44 | **hands-on-lab (25)**, thermodynamics (4), internet/web-based-learning (9), physical-chemistry (4), interdisciplinary/multidisciplinay (4), chemical-engineering (3) | 21.18 | 8.56 | 8 | 35 |
| | C45 | **stem (22)**, inquiry-learning (6), xapi (5), personalisation (3), inquiry-based-learning (8), technology-enhanced-learning (3) | 6.14 | 13.01 | 11 | 33 |
| | C46 | **virtual-reality (68)**, virtual-lab (249), physics (9), experimental-platform (3), gamification (15), augmented-reality (21) | 7.21 | 3.37 | 40 | 314 |
| | C47 | **iot (64)**, robotic-arms (5), distance-lab (4), embedded-systems (9), energy-efficiency (8), fog-computing (5) | 6.9 | 2.12 | 14 | 80 |
| | C48 | **online-lab (54)**, lms (13), a/b tests (3), remote-control (5), design-of-experiments (3), control-engineering-education (13) | 3.54 | 2.83 | 11 | 78 |
| | C49 | **educational-tools (41)**, simulation (124), online-course (3) , simulation-training (5), curriculum (8), nursing (6) | 7.62 | 3.37 | 15 | 169 |
| | C50 | **e-learning (118)**, laboratory-class (3), distance-education (15), smart-devices (4), laboratory-exercise, ict (7) | 8.35 | 1.36 | 16 | 135 |
| | C51 | **simulated-lab (34)**, covid (18), system-dynamics (5), undergraduate (5), human-computer-interaction (7) | 2.52 | 1.41 | 7 | 61 |
| | C52 | **labview (19)**, power-electronics (5), virtual-instrument (14), real-time (4) | 0.73 | 3.21 | 7 | 35 |
| | C53 | **digital-lab (4)**, problem-based-learning (10), active-learning (16) | 0.12 | 6.25 | 2 | 26 |

**C**: **C**entrality **D**: **D**ensity **CD**: **C**ore **D**ocuments **UD**: **U**nion **D**ocuments each case all the elements included in each cluster and the number of documents found with the software tool.

Table 2: Clusters of relevant keywords, two-year periods.

The table also contains the values for centrality and density as well as the number of core documents, i.e. documents with at least two keywords that are included in the thematic network. To obtain the total number of appearances, all secondary documents, i.e. the documents with a single keyword included in the thematic network, are added [48].

Figures 9 and 10 show the development of the keywords REMOTE LAB (clusters, C3, C11, C22, C33 and C44) and virtual lab (C3, C14, C23, C35 and C46) over the five biennia of the second decade studied, which evolve in different clusters.

Over the last analyzed biennia, there has been a strong push towards federation and consortium building, as well as a continuing trend towards the use of RLMS systems to remotely access and control physical lab equipment for experimentation.

o

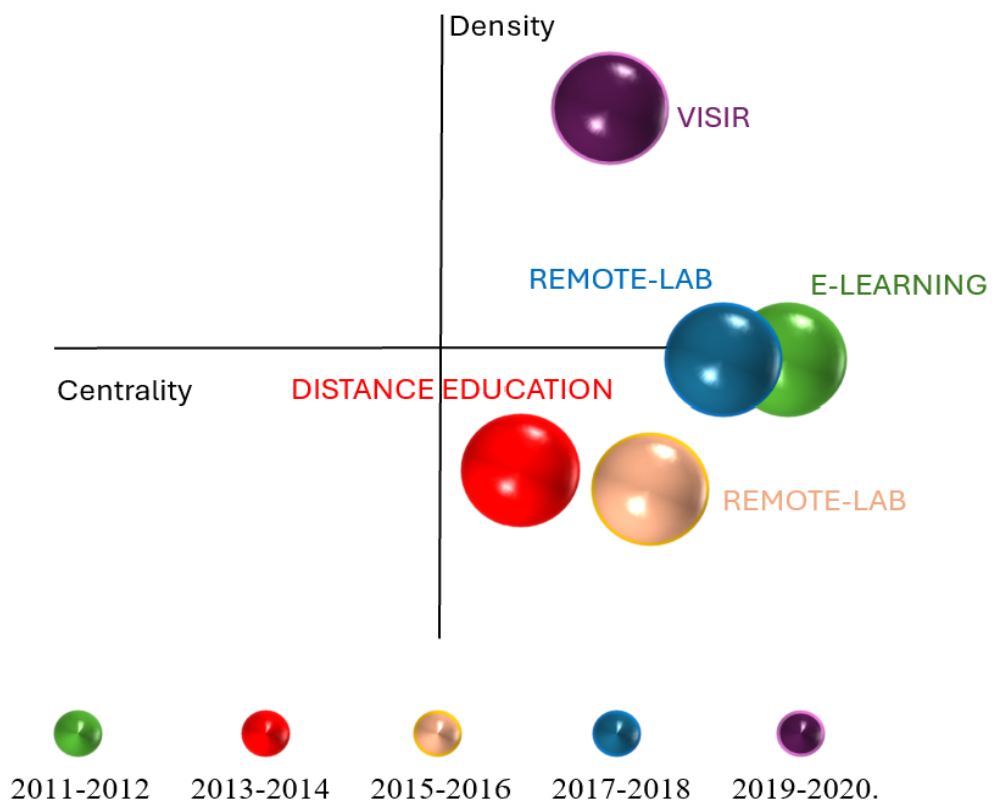

Fig. 9: REMOTE LAB Keyword evolution.

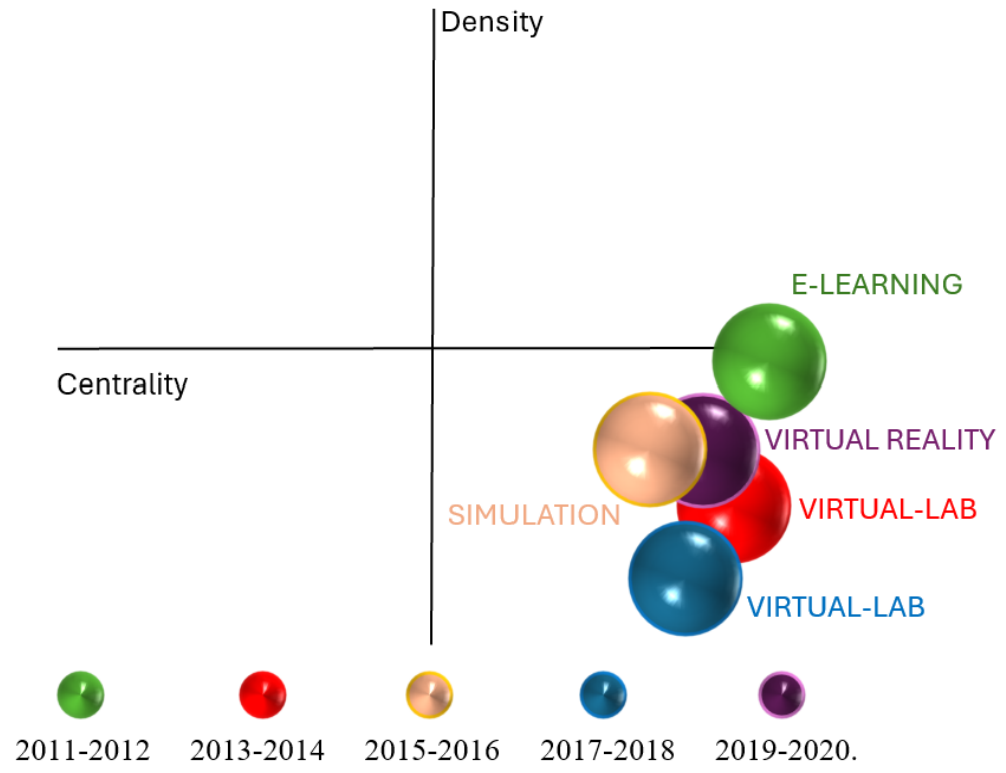

Fig. 10: VIRTUAL LAB Keyword evolution.

The emergence of new technologies linked to the development of Industry 4.0, IoT, or Internet of Things, digitalization, as well as virtual and augmented reality is observed, which can provide resources to revolutionize science education by providing students with access to high quality, immersive and interactive learning experiences.

In the next section of this paper, the observed trends and avenues for future work will be detailed in greater depth.

## VI. Analysis

The results of the systematic literature review show that virtual and remote laboratories have followed different evolutions during the decade 2001-2010 with both alternatives presenting similar weights and presence in the literature. From the second decade, the studied biennia show a rapid acceleration of remote laboratories in becoming a driving theme, while virtual laboratories have also evolved, but at a slower pace.

In the case of virtual laboratories, in the different periods they have belonged to different thematic clusters, going from VIRTUAL LAB to SIMULATION in the second decade, becoming a transversal tool with strong participation in simulated experimentation. In the case of remote laboratories, it forms a thematic cluster with great centrality and with strong participation in the formation of laboratory consortia and federations.

Analyzing the years 2019 and 2020, remote laboratories have positioned as a central tool when developing non-traditional remote experimentation, with the VISIR project standing out as the one with the largest presence at the European level, with some South American countries such as Brazil and Argentina also installing their equipment with European funding within the VISIR and VISIR + projects. Other countries have also joined such as Costa Rica at the UNED University, the University of Georgia in the United States and Morocco University Hassan.

The sustained advance of the Internet of Things (IOT) [76], [141] has also favored the growth and use of the Internet for the digitization of information, both in the academic and industrial world. With the emergence of this fourth industrial revolution, the trend to device diversity and connectivity for data collection and automation of low-complexity routine tasks helps enable advances in learning technologies.

Other highly important topics related to digitization are: BIG DATA, ARTIFICIAL INTELLIGENCE, MACHINE LEARNING, NEURAL NETWORKS, AUGMENTED REALITY, among others related to Industry 4.0. These have emerged since 2012, with a strong growth and presence onward from the 2015-2016 biennium. These topics have a high content and value around the development of software tools, algorithms, and real-time processing of a large amount of global data to favor and inform decision-making, topics that are driving forces in current research and surely in the next years.

ROBOTICS is another area that, together with AUTOMATION, has allowed the optimization and improvement of both industrial production and academic training and training proposals when it comes to bringing the use, handling, and programming of high-cost and sophisticated equipment closer to users and students. Likewise, they are terms closely related to Industry 4.0.

The research questions introduced at the beginning of this research paper are now answered.

*RQ1 What are the laboratory scenarios used in education? And what are the scope delimitations of each scenario?*

In section IV we analyzed laboratory scenarios developed at educational institutions to provide practical competencies based on controlled scenarios. Derived from this analysis, together with the SCIMAT results of Section V, scenario development has focused on real systems for distributed users, fostering anywhere and anytime scenarios. These laboratories can be either centralized or distributed. Analysis is reflected in our new taxonomy, shown in Figure 11. This taxonomy makes concrete what twenty years of research work has produced.

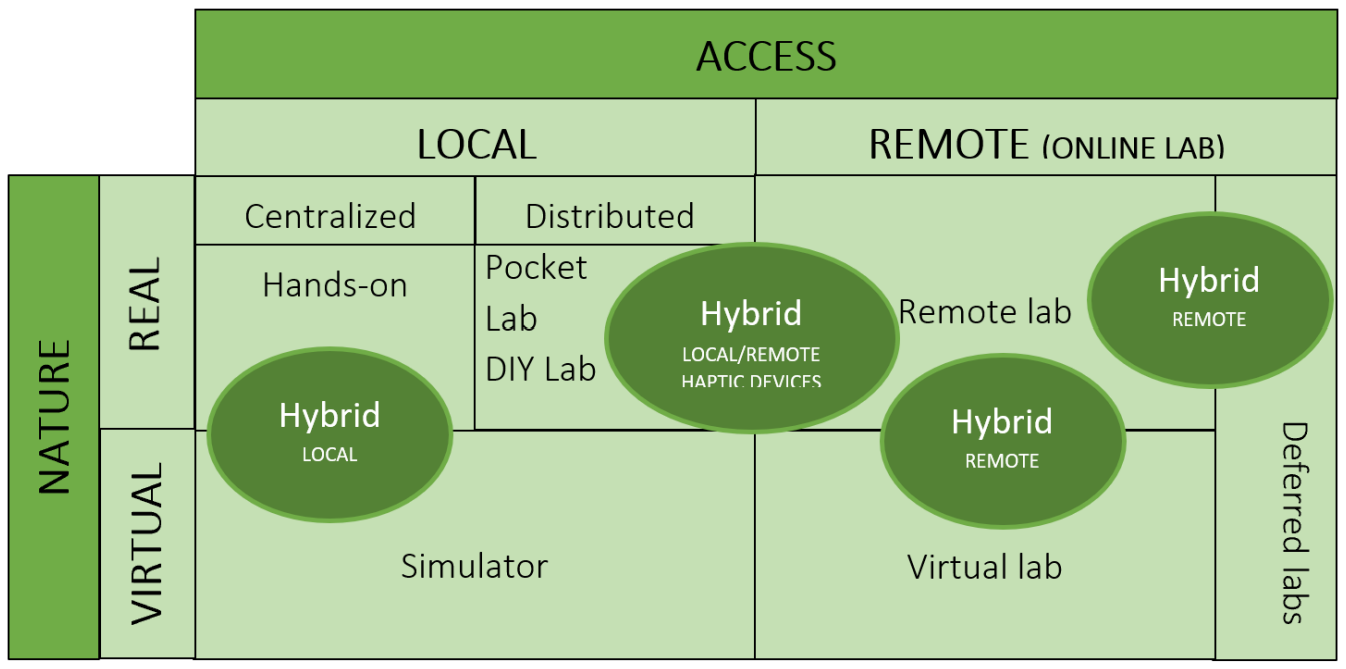

Fig. 11. A Taxonomy of Lab Work

In those laboratories, where access or interaction occurs locally, the nature of the experiment can be real or simulated. In a real system, scenarios can be defined where the laboratory can

o

be centralized or distributed. In both cases, the interaction between user and system can be computer-mediated, for example by means of Virtual Instruments. The characterization of the scenario is based on the relative location and nature of the laboratory, and not on the type of the laboratory interface. Within this scenario, one can identify centralized laboratories "hands-on labs" defined as traditional laboratories located in educational institutions, generally of high cost and with restrictive timetables; and distributed laboratories "pocket and DIY labs" defined as compact and portable systems, generally of low or moderate cost, formed by sensors and actuators, and designed for scientific and educational experimentation anywhere and anytime. The DIY, pocket or FabLabs are lab scenarios adapted to the changes promoted in student-centered educational paradigms.

Distributed laboratories based on simulated systems are "lab simulators" that are software-based, generally commercial, whose interface simulates, to some degree, a real system/laboratory, either in the modeled system or in the equipment and instrument panels. In other words, we can distinguish between simulators whose interface is a symbolic representation, and simulators whose interface virtualizes a real system. Therefore, the authors of this article do not consider as lab simulators those which are based on symbolic representations, since their purpose is to work as a calculation instrument, and not as a tool to provide experimental or practical capabilities.

Remotely accessible labs based on real systems are "remote labs" on which the scientific community has focused the greatest effort and development. These efforts for the provision of remotely accessible real scenarios to different learning environments has encouraged the application of different approaches in the system architecture to suit educational needs [7,72,111,159]. Thus, different types of remote laboratories can be distinguished grounded on the type of user-laboratory interaction and the type of generated service. This classification, together with an example of each modality, is shown in Figure 12, while more examples are given in section IV.

| REMOTE LABS | | TYPE OF INTERACTION | | |
|---|---|---|---|---|
| | | BATCH LAB | SYNCRONOUS LAB | OBSERVATION LAB |
| TYPE OF SERVICE | CONCURRENT LAB | Visir [139] | Arduino Robot [41] | Meteolab [112] |
| | MONOUSER LAB | FPAA lab [174] | 3-axis-Portal [147] | |

Fig. 12. Remote labs scenarios.

Regarding the type of interaction between the user and the laboratory, we can distinguish three clearly different types: batch, synchronous and observation laboratories.

*Batch laboratories* are based on communication protocols with experiment configuration request queues [30,69]. Here, the user configures the experiment and sends this configuration to the remote laboratory for experimentation. This type of remote lab is typical of systems where a program, routine or algorithm needs to be loaded on a device [39,146,174]. The service type of batch labs can be either mono user or concurrent. As an example of a batch laboratory, we can mention the case of the VISIR laboratory architecture, which is based on a type of requests/responses interaction. It allows concurrent users, since the response of the electrical/electronic circuits is very fast, in the order of milliseconds, which gives the feeling of private control of the laboratory.

*Synchronous laboratories* are those in which the laboratory-user interaction is like that carried out in hands-on labs, responding in real time to the modifications made by the user. Obviously, this type of interaction in remote laboratories only allows concurrent access through hardware replication or the federation of the laboratory.

*Observation labs* are those in which the user cannot change anything in system. These are laboratories dedicated mainly to monitor and obtain historical data [211].

The authors of this article do not consider the possibility of a mono user observation lab, since limiting its access to a single user would be an error of bad design (since the observation lab must be concurrent by nature).

Regarding hybrid laboratories, these are alternatives for very specific applications, where it is required to enrich a real environment with some kind of simulation.

*RQ2 What are the types of emerging labs scenarios?*

Great efforts have been made to provide real-world scenarios for distributed users. As shown in the results section, these efforts were particularly relevant to the increasing importance of remote labs in the field of experimentation. This has been aided by the development of technology, the emergence of open-source solutions and the promotion of low-cost hardware. Another factor favoring growth is the plasticity and agility of non-traditional labs in incorporating new technologies adapted to the educational environment. This fact was demonstrated in the results section in examples such as the emergence of Industry 4.0 [76], [54], [81]. In line with the do-it-yourself paradigm, this has also enabled the emergence of mobile laboratory alternatives that provide students with new experimentation scenarios. As a result, a variety of new and innovative ad hoc labs have emerged for the hands-on learning process (refer to Figure 11). Alternative labs based on real-world systems, such as pocket labs, time-shifted labs or hybrid options, have increased real-world experimentation for distributed users.

These new solutions allow adapting laboratory experiences to educational processes. In this way, the laboratory is integrated, developing from being an isolated activity to becoming a part of the whole learning process.

Deferred labs are an interesting alternative to provide real (pre-recorded) experiences remotely, which can be easily accessed via the internet. They make it possible to offer real or quasi-real experiences on a large scale. However, their potential based on ultra-concurrent access and stability, without maintenance or breakdown, make them an ideal, if not the best, solution for massive environments such as MOOCs or courses with a high number of students or laboratory experiences with recurring access.

Regarding hybrid solutions, where characteristics of real and simulated laboratories are combined, there have emerged various alternatives, that have complemented the academic offer, especially for types of laboratories with non-repetitive tests and experiences.

Virtual laboratories have developed as a further, in many cases complementary, tool. These new systems are mainly focused on areas where another type of remote solution is not technologically possible, such as in health education, or where they are very expensive systems to implement in a remote laboratory, or in K-12 educational environments because they are reliable systems where simple and attractive interfaces can be easily created [127].

In summary, there is a strong growth of real experimentation environments for distributed users. Emerging labs scenarios are adapted to the changes promoted in the student-centered educational paradigms. In this way, the student does not only consume what the teacher "puts on the table" but the system encourages the student to be highly involved and motivated.

*RQ3 How have lab scenarios evolved in 2011-2020 decade?*

The emergence of the Internet, the advances in connectivity and the appearance of low-cost hardware have allowed the creation of technological alternatives for the collection of information in real time through Internet-connected devices.

As regards the use of the LMS (Learning Management System), in the period 2001-2010, there were 72 appearances in the study, presenting a high impact transversal behavior. In the second decade, as shown in Figure 6, encompassing the period 2011-2020, the appearance of RLMS (Remote Learning Management System) gained strength with the growth and development of remote learning platforms.

The RLMS platforms made it possible to create systems where the access and control of remote laboratories is centralized in terms of authentication, authorization, scheduling, and lab resources management, becoming a transversal tool for the operation of remote laboratories.

There is also a growing integration of laboratory networks to share resources, foster development in a specific area of expertise and promote research.

This collaborative linkage has allowed enriching the academic and research offerings, where several educational institutions in different countries and regions participate to provide a high availability of academic learning resources.

One aspect that has gained relevance is the provision of laboratory experiences based on real systems in massive environments, MOOCS and MOOLS, where educational institutions have been able to cover a wide range of topics (not only in science) by offering content in a variety of formats, open to the community.

Another important trend is the creation of the laboratory exploitation scenarios, including immersive interaction (e.g., 3D, augmented reality) [118], [122], enriching the user's experimentation process, which will be a strong trend in the coming years in laboratory development.

Another case to be mentioned is LabVIEW, a widely used tool since the emergence of remote laboratories as a hardware and software alternative. Nevertheless, in recent years, its use has decreased with the growth of low-cost hardware platforms. The concept of collaborative laboratories [115], [143], [150], [211] is gaining prominence to make up for the shortcomings of online laboratories, in terms of teamwork capabilities. This is an important challenge for the evolution of remote laboratories in the future.

## VII. Conclusions

Although this highly-specialized community does have a common framework in this area, this work shows that there is still a certain conceptual ambiguity of concepts, in terms of the classification of laboratories found in its literature. This paper contributes to the literature by providing a framework to delimit and define each term and scenario, offering a common language when talking about experimentation scenarios.

Over the last few years, new pedagogical tools have emerged, based on the strong impact of the Internet and the access to technology, where remote laboratories have become an important way to strengthen the development of practical activities, putting the student at the center of the learning process.

There is a strong trend towards laboratory federation through RLMSs. The creation of collaborative consortia between academic institutions links different regions and countries to unify learning criteria, methodologies and join efforts for better disruptive pedagogical alternatives in the age of knowledge and information. Therefore, federation protocols foster synergies between institutions and countries.

It is important to highlight the potential future contribution of the IEEE 1876 Standard [212], which establishes certain guidelines for the design, implementation, and general operation of remote laboratories, thus promoting collaboration and innovation. However, the use of the architecture proposed by this standard faces some resistance because the framework of many of the laboratories must be redesigned to comply with the standard, and for this reason they are not currently being implemented.

One future direction for the standard would be a new version that covers the full range of scenarios outlined in Figure 11.

The development of new pedagogical tools in the field of experimentation is oriented towards the implementation of solutions based on real systems, deferred laboratories, federations and at least some degree of standardization either by the IEEE 1876 Standard or by RLMS.

As aforementioned, there have been discontinued efforts to provide collaborative online laboratory scenarios, which have not been successful due to an approach based on ad-hoc solutions. In this sense, an environment through an RLMS that universalizes this service is an interesting line of research to cover this gap in online scenarios. The possibility of alternating between collaborative and individual environments would make it possible to obtain the capabilities of both scenarios.

There are alternatives through private companies such as Labsland [39] that provide, among other alternatives, online laboratories for education. This service model promotes several aspects discussed in this paper through the trends analyzed: federation, RLMS platforms with standardized communication protocols that allow easy integration into other platforms, promotion of deferred laboratories and redundancy in remote laboratory services.

The challenge for the future of experimentation seems to be to increase the pedagogical offerings and the sustainability of

o

parallel collaborative spaces that may promote inclusion and universal access to simple technological experiments that may enhance the learning experience of students worldwide. For now, laboratories based on real systems are positioned as the main tool for the acquisition of practical skills, focusing efforts on remote labs solutions, while virtual laboratories serve as additional tools, especially in areas where other remote solutions are not feasible.

From an availability and reliability service point of view, Deferred Labs are a valid option when it comes to providing a reliable educational tool that minimizes the possibilities of failure, with the main advantage of being able to offer a real mass access experience, and low-cost maintenance. These labs have a potential application to shift specific lab experiences or complete remote labs and hands-on labs to deferred labs. In this way, maintaining the educational laboratory resource/tool indefinitely, so that no breakdown or failure would cause a discontinuity or even cancellation of service provision.

## ACKNOWLEDGMENT

The authors thank the Escuela de Doctorado de la UNED, Spain, for its support. This work was also supported by the project ECoVEM—European Centre of Vocational Excellence in Microelectronics, nº. 620101-EPP-1-2020-1-BG-EPPKA3-VET-COVE. Also part of the activities of the UNED Research Group I4Labs, Industria Conectada y Tecnologías Educativas para la Ingeniería, and Innovative Educational Group, G-TEIC, Tecnologías Educativas para la Industria Conectada.

o

o

o

o

o

o